\documentclass[10pt,twocolumn,letterpaper]{article}

\usepackage{cvm}
\usepackage{times}
\usepackage{epsfig}
\usepackage{graphicx}
\usepackage{amsmath}
\usepackage{amssymb}
\usepackage{amsmath,mathtools}
\usepackage{caption}
\usepackage{subcaption}
\usepackage{flushend}
\usepackage{verbatim}
\usepackage{multirow}
\usepackage{rotating}
\usepackage{tabu}                      
\usepackage{booktabs}

\usepackage[pagebackref=true,breaklinks=true,letterpaper=true,colorlinks,bookmarks=false]{hyperref}

\cvmfinalcopy

\def\cvmPaperID{****} 

\ifcvmfinal\fi
\begin{document}

\title{Animation-by-Demonstration Computer Puppetry Authoring Framework}

\author{
Yaoyuan Cui\\
{\tt\small yaoyuan.cui@siu.edu}
\and
Christos Mousas\\
{\tt\small christos@cs.siu.edu}
\and
Graphics \& Entertainment Technology Lab\\Southern Illinois University, Carbondale, IL 62901, USA
}

\maketitle

\begin{abstract}
This paper presents Master of Puppets (MOP), an animation-by-demonstration framework that allows users to control the motion of virtual characters (puppets) in real time. In the first step, the user is asked to perform the necessary actions that correspond to the character's motions. The user's actions are recorded, and a hidden Markov model (HMM) is used to learn the temporal profile of the actions. During the runtime of the framework, the user controls the motions of the virtual character based on the specified activities. The advantage of the MOP framework is that it recognizes and follows the progress of the user's actions in real time. Based on the forward algorithm, the method predicts the evolution of the user's actions, which corresponds to the evolution of the character's motion. This method treats characters as puppets that can perform only one motion at a time. This means that combinations of motion segments (motion synthesis), as well as the interpolation of individual motion sequences, are not provided as functionalities. By implementing the framework and presenting several computer puppetry scenarios, its efficiency and flexibility in animating virtual characters is demonstrated. 
\end{abstract}


\section{Introduction}
\label{sec1}
The recent increase in affordable motion capture and sensing interfaces (i.e., Kincet, Wiimote, and Leap) permits users to interact with displayed and virtual content in more complex and advanced ways, thus allowing them to become more immersed in the virtual experience. Such devices can capture the full-body's movements or simply finger gestures or facial expressions. Based on previously published work \cite{ref37}, it can be said that the ability to interact with a virtual environment through body movements enhances the user experience.

Such interfaces, have been used to control virtual characters as well as to allow users to navigate and interact with virtual environments. So-called computer puppetry or performance animation methods are mainly used for controlling characters, and the entertainment, gaming, and virtual reality industries generally benefit from them. The main disadvantage of most performance-driven character control approaches is their limited ability to work only with human-like virtual characters. However, in computer puppetry, users should be able to interact with a variety of characters and creatures, even those that are not humanoid, based on body movements, finger gestures, and facial expressions.

In the skeletal-based, performance-driven animation of humanoid characters, the user's joints are mapped to the joints of a virtual character. However, in some cases, the target character may not be a humanoid. For example, it might be a virtual creature or an object, such as a table or a chair, that needs to be animated. When such complex non-human characters and objects need to be animated, it is necessary to define advanced mapping techniques, which are generally called retargeting \cite{ref26}. It should be noted that retargeting techniques are responsible for transferring the motion that belongs to a particular morphology to a different one. The disadvantage of such retargeting techniques is mainly the time-consuming manual work that is required to define the correspondence between the user's poses and the non-human character's poses.

This paper focuses on the user's action-based control of a virtual character's motion. Specifically, it addresses the issue of controlling an individual character's independent motion by a particular action performed by a user. Such character control mechanism can also be described as a highly constrained computer puppetry control method. This means that the virtual puppet is able to perform only specific motions as well as one motion at a time, and the performers can use their bodies or body parts (e.g., fingers and face) to control the motions independently. To solve this constrained problem, a hidden Markov model (HMM) that learns the temporal profile of the user's actions is used. During the runtime of the applications, the real-time prediction is achieved using the forward algorithm, which predicts the action of the user and consequently the motion of the character that should be displayed as well as the evolution of the character's motion based on the evolution of the user's actions.

In the Master of Puppets (MOP) framework, it is not necessary to perform exhaustive mapping between the user's body parts and the character's motions. This is achieved by directly associating the time progression of the user's action with the time progression of the character's motion. Therefore, a generalized soft mapping process is used in this approach compared to other methodologies that highly constrain this process. The generalization process for controlling virtual characters makes it possible for the user to independently control the motions of different characters (humanoid or non-humanoid), even if the target characters are dissimilar or have different morphological variations compared with the user. To summarize, the MOP framework can be characterized as follows:
\begin{itemize}
\item	\textbf{Easy to Use:} Users are not required to manually define the correspondence between their joints and the character's joint angles or control points. Each user captures their actions and assigns them to the corresponding animation of the virtual character.
\item \textbf{Fast:} The framework learns the temporal profile of each action only a few seconds after capturing the actions of the user.
\item \textbf{Flexible:} The framework provides the ability to animate the target characters without the need for morphological similarities to a human and without the need to use only animations of skeletal-based characters.
\end{itemize}

This remainder of this paper is organized as follows. Section \ref{sec2} discusses previous work that is related to the proposed methodology. Section \ref{sec3} presents the core part of the MOP framework, and Section \ref{sec4} discusses the implementation details, examples, and evaluation of the presented framework. Finally, Section \ref{sec5} provides conclusions and possibilities for future work.

\section{Related Work}
\label{sec2}
Different research fields can include the MOP framework. Therefore, this section outlines previous work related to MOP and discusses the contributions of the presented framework.

\subsection{Input Devices}
\label{sec21}
Interactive character control mechanisms can be classified based on the devices used to animate virtual characters \cite{ref37}. The keyboard and mouse are the most common ways to control the motions of virtual characters, as well as joysticks, which are used in most game consoles \cite{ref38}. It is also important to note that more specialized control mechanisms have been introduced to animate virtual characters, such as text input \cite{ref2}\cite{ref7}, speech-based \cite{ref11}, and sketch-based interfaces \cite{ref40}; RGB-D \cite{ref18}\cite{ref19} and IR \cite{ref6} sensors; data gloves \cite{ref12} or color gloves \cite{ref13}; simple accelerometers \cite{ref14}\cite{ref17}; and even 3D-printed custom interfaces \cite{ref41}. MOP allows the use of different motion sensing devices to control a virtual character. In its current version, the MOP framework supports the use of Microsoft's Kinect and Leap motion devices. The flexibility of handling input from more than one device simultaneously is another unique feature of MOP. (Section \ref{sec412} and the video that accompanies this paper present an example in which more than one device is used simultaneously for capturing the user's activity.)

\subsection{Character Animation}
\label{sec22}
Characters can be animated in several different ways, such as based on the keyframing method, which is characterized by its time-consuming process as well as by the specialization the animator should have. Computational methods use existing motion data (data-driven methods) and adapt the motion to fulfill the necessary constraints of the virtual environment. Among the numerous data-driven motion synthesis techniques that exist, interpolation \cite{ref21}, blending \cite{ref22}, splicing \cite{ref24}\cite{ref1}, and warping \cite{ref25} are used most extensively to make existing motion data reusable. A variety of statistical and dimensionality reduction methods have been used in the past to handle data efficiently. Examples include principal component analysis (PCA) \cite{ref23}, Gaussian process latent variable models (GPLVM), kernel methods \cite{ref8}\cite{ref15}, and multi-dimensional scaling \cite{ref44} methods. Such methods have been used to parameterize the existing motion data to animate virtual characters in a proper way \cite{ref43}. MOP can be characterized as a data-driven character animation framework, since the characters are animated using motion data. However, MOP does not consider any motion synthesis technique in its current form. It only provides users the ability to control the evolution of motions that are assigned to virtual characters.

\subsection{Performance Animation}
\label{sec23}
In performance animation, users control a virtual character with their bodies. Motion capture systems usually provide the necessary input signals retrieved either from accelerometers \cite{ref28} or from optical input \cite{ref30}. Then, techniques based on kinematics-related solutions \cite{ref32} or data-driven motion reconstruction \cite{ref34}\cite{ref4} are used to ensure that the reconstructed motion lies in the pose space of the existing motion data. It should be noted that data-driven techniques can reconstruct natural-looking poses using a very small number of control inputs. The reduction of the error is achieved due to the prior knowledge that is obtained from the example data. The use of one \cite{ref10}, two \cite{ref36}, and six \cite{ref34} input(s) has also been examined. It should be noted that statistical models based on large motion datasets \cite{ref30}\cite{ref34} are used to build prior knowledge to achieve such an under-constrained reconstruction process. However, even if the use of one \cite{ref10} or two \cite{ref36} input(s) provides a realistic motion, complex motions cannot be reconstructed, especially when the user performs an action with a body part that is not captured. MOP relates to performance animation since users must use their bodies or body parts to control the motion of virtual characters.

\begin{figure*}[htb]
\centering
\includegraphics[width=1\textwidth]{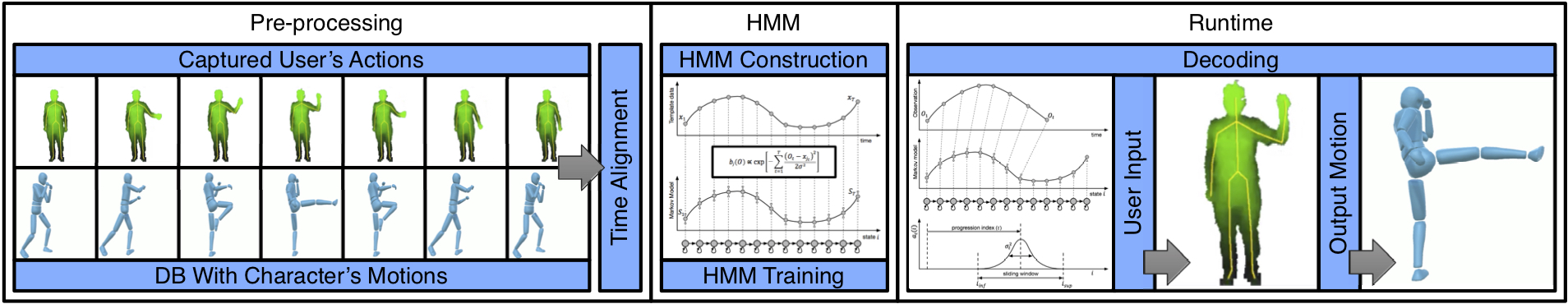}
\caption{The architecture of MOP framework.}
\label{fig1}
\end{figure*}

\subsection{Computer Puppetry}
\label{sec24}
In computer puppetry, users control the motion of a virtual character not necessarily by defining a joints-to-joints mapping. Mid-air gestures and machine learning techniques have also been used to define correspondence between the input actions of a user and the output motions of a character. In most computer puppetry methods, it is quite important to choose the right input device \cite{ref45}. Dontcheva et al. \cite{ref46} proposed a method that does not require the time-consuming manual selection of a character's body parts. Instead, this is achieved by aligning the temporal sequences using canonical correlation analysis. Seol et al. \cite{ref47} proposed a computer puppetry approach based on a correlation mapping process of the features of a user's actions and a character's motions. The approach developed by Chen et al. \cite{ref19} maps joint positions between the user and the object. The animation is achieved using mesh deformation techniques. Vögele et al. \cite{ref20} developed a method in which the motions of two users are mapped in a variety of quadrupedal characters. Yamane et al. \cite{ref49} proposed a method that works offline and performs mapping between human motion and non-human motion. To achieve natural-looking results, the user should manually define 30 to 50 correspondences. Finally, hybrid controllers \cite{ref20}\cite{ref31} that allow the user to navigate and interact with virtual objects in virtual environments in a more sophisticated way through mid-air gestures and the direct manipulation of body parts have been also introduced. MOP differs from previously developed puppetry approaches, since users can control the time progression of the motion instead of the motion itself, and the hard-constrained piece-by-piece mapping process is not applicable.

\subsection{Score Following}
\label{sec25}
The MOP framework is inspired by existing work that examines ``score following" or ``gesture following". Such methods are responsible for learning the temporal profile of a continuous stream of data. Various applications benefit from score following, since such methods help improve the way users interact with computers. Real-time music accompaniment \cite{ref52} and the gestural control of sound \cite{ref33} are good examples of this. The most common approaches for score following include the use of dynamic time warping (DTW) \cite{ref39}, HMM \cite{ref50}, neural networks (NN) \cite{ref48}, support vector machines (SVM) \cite{ref29}, and dynamic programming (DP) \cite{ref27}. The score following method developed in this paper is based on HMM.

\subsection{Contribution}
\label{sec26}
The MOP framework contributes several unique features. First, the correspondence between a user's actions and a character's motions are built automatically even if the input parameters are not directly related to the target parameters. Second, the HMM allows users to control the evolution of a character's motion based on their own activities. Given this, the output motion is not synthesized but rather displayed just as it is. Third, the flexibility of the proposed method to handle different characters as well as a variety of inputs are two additional important advantages. MOP automatically handles the data and allows the user to interact immediately with the virtual character. The examples in this paper show that without any additional effort, MOP provides users the ability to control more than one character simultaneously, and it allows more than one user to control a target character. It is assumed that the contributions provided by the proposed method could benefit the entertainment community, since all the functionalities implemented within a single framework can be used for a variety of computer puppetry interaction scenarios. For example, data-driven digital shadow puppets, which are a good example of animating constrained virtual puppets using specific actions and corresponding output motions, could benefit from MOP.

\section{Methodology}
\label{sec3}
This section presents the key components of the MOP framework. Specifically, it describes how users are able to control virtual characters based on their actions. The architecture of the MOP framework, which is divided into three parts (pre-processing, HMM construction and training, and runtime character control) is illustrated in Figure \ref{fig1}.

\subsection{Representation of Motion Data}
\label{sec31}
MOP requires at least two motion sequences. The first one is the user's action, and the second is the motion that animates the virtual character. The first motion is captured using one of the compatible sensors. The user designs the second set, or if a humanoid character is being animated, it can be captured. The rest of this section describes how both types of data are represented.

\subsubsection{User} 
\label{sec311}
The user's action is represented as $X_{1:T}=[x_1,...,x_T ]$, where $T$ denotes the total number of frames. Each $x_t$ contains the pose of the user at time $t$, which can also be represented as rotations of joint angles $r$, such as $x_t=[ r_t (1),...,r_t (L)]$, where $x_t \in \mathbb{R}^{dx}$ and $L$ denote the number of captured joints of the user. It should be noted that the captured position of the user's root is not included. It is assumed that the virtual character will remain in the same position in the virtual environment. The same representation is used when capturing either the full-body motion or the finger motion of a user. However, the input parameters can be also retrieved from the face of the user. In this case, the user's motion is represented by control vectors $x_t=[p_t (1),...,p_t (G)]$ where $G$ denotes the total number of captured features of the user's face at time $t$. It should be noted that each $p_t (g)$ is represented based on its local position according to the head root.

\subsubsection{Virtual Character} 
\label{sec312}
The ways in which a virtual character are animated are based either on skeletal or blendshape-based keyframes. In both cases, a character's motion can be represented as $Y_{1:T}=[y_1,...,y_T ]$. As before, $T$ denotes the total number of frames (both $X$ and $Y$ should have equal lengths; therefore, a simple time alignment process regularizes the corresponding data). Each $y_t$ contains the pose of a character, which can also be defined as the rotation, $q$, of a joint angle, such as $y_t=[q_t (1),...,q_t (K)]$, where $y_t \in \mathbb{R}^{dy}$ and $K$ denotes the total number of joints of a virtual character. When the motion of a virtual character is represented by blendshapes, a frame of the animated sequences is defined as $y_t=[g_t (1),...,g_t (F)]$, where $g_t (f)$ denotes the evolution (the weights) of the $f-th$ blendshape at the $t-th$ frame.

\subsubsection{Time Alignment} 
\label{sec313}
In the MOP framework, the user is asked to perform actions that are later used to train the HMM. To properly provide the activity following functionality, $X$ and $Y$ should always have the same duration, $T$. Therefore, right after capturing the actions of a user for particular motions and before training the HMM, normalization of the corresponding durations is used to make sure both $X$ and $Y$ have the same lengths. Thus, a soft mapping between $x_t$ and $y_t$ is achieved.

\subsection{HMM-Based Action Follower Mechanism}
\label{sec32}
This section presents the HMM, which allows the user to control a virtual character based on its body activities. Two steps are presented. In the first step (training process), an HMM is trained to learn the parameters, $\lambda$, of the model. The parameters of the HMM are defined as $\lambda = \{ a_{ij}, \pi_i, b_i \}$, where $a_{ij}$ represents the state transition matrix and denotes the probability of making a horizontal transition from the $i-th$ to the $j-th$ state, $\pi_i$ represents the prior vector and denotes the initial distribution vector over the sub-states of the model, and finally, $b_i$ represents the observation probability distribution and indicates the probability of the production state. In the second step (decoding process), a user's activity is predicted in real-time. The HMM allows the user to animate a virtual character by controlling its motions independently based on different body activities. According to Berndt and Clifford \cite{ref42}, the developed model can be considered as a hybrid approach between HMM and DTW.

\subsubsection{Training Process}
\label{sec321}
The training process is based on setting an HMM to recognize the captured motion of the user. Since this approach does not consider capturing variations of a user's action, a single action is used to animate a single motion of a character. Thus, someone can easily realize that limited training data are available. A template-based DTW method is used, and a single example of the captured data is set for the learning process. The structure of the state of the HMM is set directly from a single template, which is a time sequence that is represented by $x_1,...,x_T$ where $x_t$ represents the posture of a user, as described in the previous section. 

\begin{figure}
\centering
\includegraphics[width=1\columnwidth]{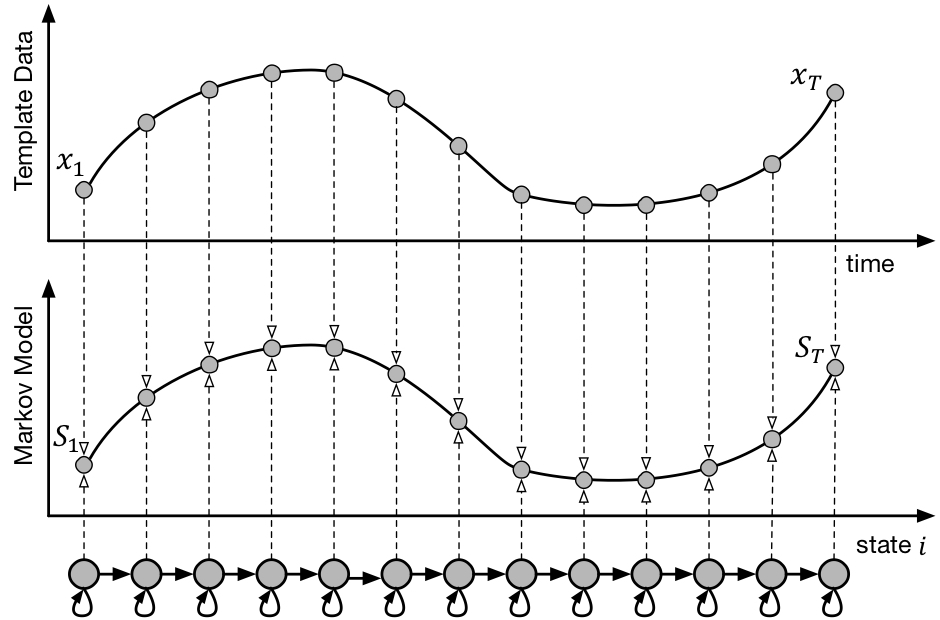}
\caption{A Markov chain $S_1,...,S_T$ is generated based on the captured motion sequence $x_1,...,x_T$ (template data) during the training process of the HMM.}
\label{fig2}
\end{figure}

The $x_1,...,x_T$ sequence is then used to create a left-to-right Markov chain that is represented by $S_1,...,S_T$ states. Only two transitions are considered in this implementation: the self transition, $a_{ii}$, and the next transition, $a_{ij}$. The $a_{ij}$ transition describes the state transition probability distribution between state $i$ and $i+1$. In this case, it is also considered that the data are regularly sampled. Therefore, the transition probabilities are set manually as $a_{ii} = a_{ij} = 0.5$, where a value of $0.5$ corresponds to an average transition time that is equivalent to the original sequence. Figure \ref{fig2} illustrates the training process of the developed model and, more specifically, the ways the $S_1,...,S_T$ states that represent the left-to-right Markov chain are constructed.

Considering an observation $O$ ($O$ is a measured vector of length $T$), the probability $b_j(O)$ in a state $j$ is set to a Gaussian distribution with vector $x_j$ as the center. Apparently, a single template is not enough to fully estimate the required distributions. Thus, a simplified form for estimating the required distributions is used based on the following equation:
\begin{equation}
b_j (O) \propto \exp \Bigg[-\sum_{t=1}^T \frac{(O_t-x_{j_{t}})^2}{2 \times \sigma^2 }\Bigg]
\label{eq1}
\end{equation}
where $\sigma^2$ denotes the standard deviation between the measured and template data.

\subsubsection{Decoding Process}
\label{sec332}
During the runtime of the application, the decoding algorithm runs on a growing sequence of observations, $O_{1:t} = O_1,...,O_t$. The decoding process is responsible for estimating the probability distribution, $a_t (i)$, which denotes the probability of the observation sequence, $O_{1:t}$, as well as the state, $S_i$, at the $t-th$ time step of the given model. Then, it is possible to estimate $a_t (i)$ using the forward algorithm \cite{ref51}. Three different quantities are computed from the distribution $a_t (i)$: the likelihood, $L_t$, at time $t$ of the observation sequence, $O_{1:t}$ (see Equation \ref{eq2}), the first moment, $\mu_t$, of the normalized distribution, $a_t (i)/ L_t$ (see Equation \ref{eq3}), and the variance $\sigma_t^2$ of the normalized distribution (see Equation \ref{eq4}) as presented below.
\begin{equation}
L_t = \sum_{i=1}^N a_t (i)
\label{eq2}
\end{equation}
\begin{equation}
\mu_t= \sum_{i=1}^N \Bigg[ i \times \frac{a_t (i)}{L_t} \Bigg]
\label{eq3}
\end{equation}
\begin{equation}
\sigma_t^2 = \sum_{i=1}^N \Bigg[ (i-\mu_t)^2 \times \frac{a_t (i)}{L_t} \Bigg]
\label{eq4}
\end{equation}

The likelihood, $L_t$, is used to estimate the similarity between the observation and template sequences. The first moment, $\mu_t$, is used to predict the time progression index of the input motion. This can be described as the essential output parameter of the real-time prediction process, since it enables the real-time alignment of the observation to the template sequence. This time progression is computed by $\mu_t / f$, where $f$ denotes the sampling frequency. Finally, the variance, $\sigma_t^2$, of the normalized distribution in the $a_t (i)$ state is important, since it can be used to invalidate some of the outputs from the activity follower methodology. It should be noted that the value of the variance could be considered complementary to the likelihood value, $L_t$. It provides information on the similarity between data (a small $\sigma_t^2$ value indicates that the progression index is statistically defined properly and vice versa).

To improve the efficiency of the real-time prediction process, the forward algorithm is calculated on a sliding window as:

\begin{equation}
a_1 (i) = \pi_i \times b_i (O_1 )
\label{eq5}
\end{equation}
\begin{equation}
a_{t+1} (j) = k \times \Bigg[ \sum_{i=i_{inf}}^{i_{sup}} a_t (i) \times a_{ij} \Bigg] \times b_j (O_{t+1}) 
\label{eq6}
\end{equation}

\begin{figure}[htb]
\centering
\includegraphics[width=1\columnwidth]{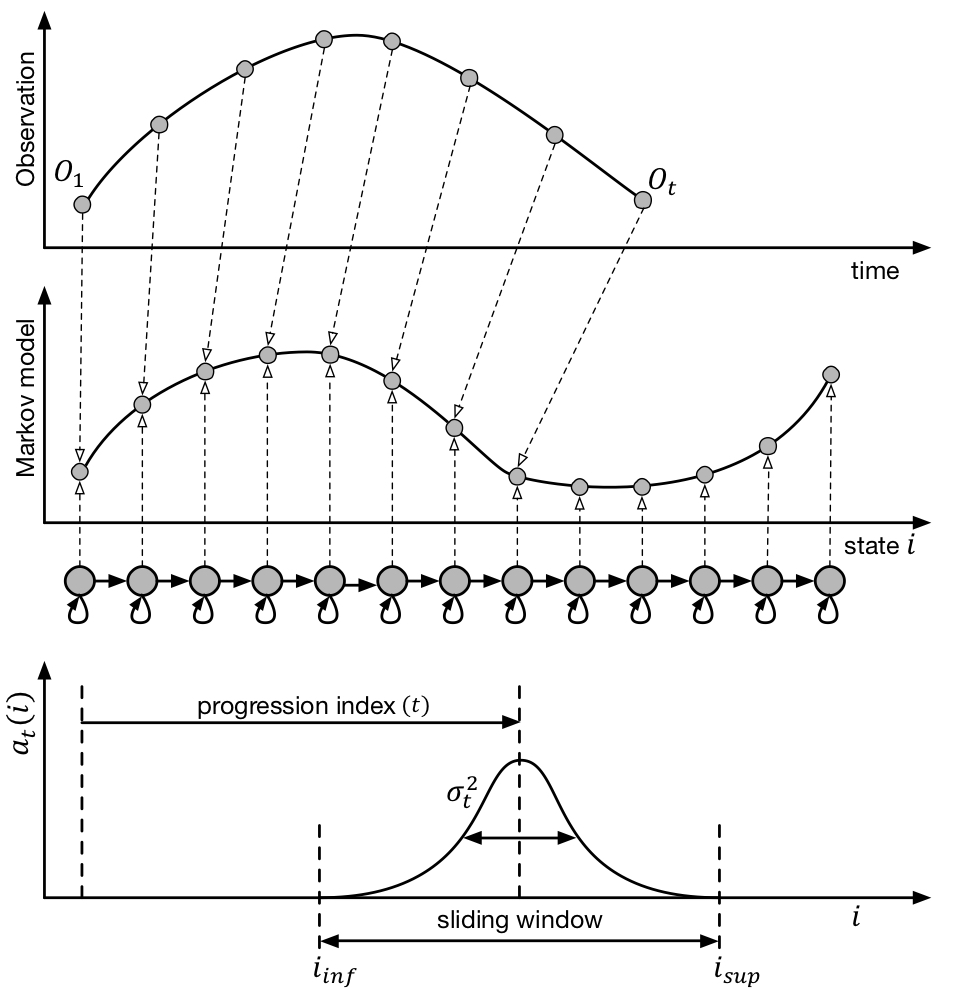}
\caption{A graphical representation of the decoding process as implemented in the MOP framework.}
\label{fig3}
\end{figure}

In Equation \ref{eq5}, $i \in [1, N]$, and in Equation \ref{eq6}, $i \in [1, T-1]$ and $j \in [1, N]$. Moreover, $\pi_i$ denotes the initial distribution in state $i$, $a_{ij}$ denotes the transition probability distribution, and $i_{inf}$ and $i_{sup}$ denote the inferior and superior indexes of a sliding windows with lengths of, $2p$ (a 20 ms window is used in the examples presented in this paper). It should be noted that the MOP user can easily adjust the window size to make the recognition process work properly. In Equation \ref{eq6}, $k$ denotes a normalization factor that results from the truncation of the sum. Generally, this normalization can be ignored if $p$ is a large number. Additionally, it should be noted that the computation of $a_{t+1} (j)$ is quite efficient, since the form of $a_{ij}$ is rather simple. Finally, $i_{inf}$ and $i_{sup}$ are generally set as functions of the index $\mu_t$ based on the following rules:

\begin{align}
\begin{split}
\text{if $(1<\mu_t \leq p)$} &
\begin{cases}
i_{inf}=1\\
i_{sup}=2 \times p+1
\end{cases}\\
\text{if $(p<\mu_t \leq N-p)$} &
\begin{cases}
i_{inf} = \mu_t-p\\
i_{sup} = \mu_t+p
\end{cases}\\
\text{if $(N-p<\mu_t \leq N)$}&
\begin{cases}
i_{inf}=N-2\\
i_{sup}=N 
\end{cases}
\end{split}
\label{eq7}
\end{align}
Concluding, Figure \ref{fig3} illustrates the decoding process presented above and shows the way the observation data are aligned with the states of the Markov model (time warping) as well as the windowing process of the associated probability function. 

\section{Implementation Details and Evaluation}
\label{sec4}
This section presents the interactive workflow of MOP as well as examples that illustrate its functionalities. MOP is also evaluated against similar frameworks/methods. 

\subsection{Implementation}
\label{sec41}
\begin{figure}[htb]
\centering
\includegraphics[width=1\columnwidth]{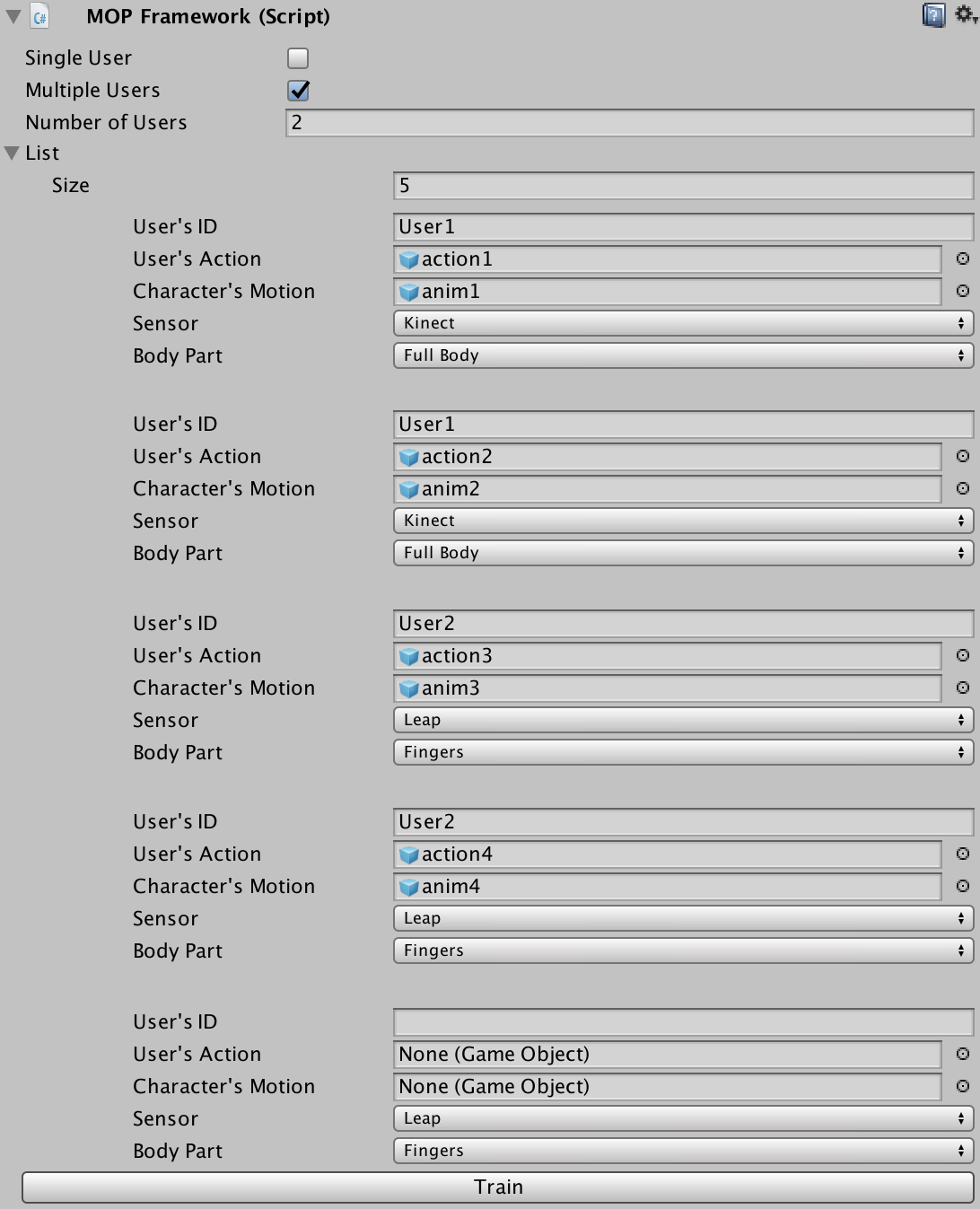}
\caption{The editor window in Unity3D that provides users the ability to author their own computer puppetry interaction scenarios.}
\label{fig4}
\end{figure}

MOP is implemented as an editor extension of the widely used game engine, the Unity3D\footnote{\url{https://unity3d.com/}}. Figure \ref{fig4} illustrates the simple interface of MOP in Unity3D. The user defines the number of action-motion pairs, assigns the captured actions to the corresponding motions of the virtual character, and finally chooses the input sources that should be used. In the current version of the MOP framework, Microsoft's Kinect\footnote{\url{https://www.xbox.com/en-US/xbox-one/accessories/kinect}} motion capture sensor with the software development kit (SDK) provided by \cite{ref5}, and the Leap\footnote{\url{https://www.leapmotion.com/}} motion controller with the SDK provided by \cite{ref3} are integrated. A downloadable version of the Unity3D package that contains the complete framework with a variety of examples can be found on the project webpage of the MOP framework.

\subsubsection{Interactive Workflow}
\label{sec411}

\begin{figure}[htb]
\centering
\includegraphics[width=1\columnwidth]{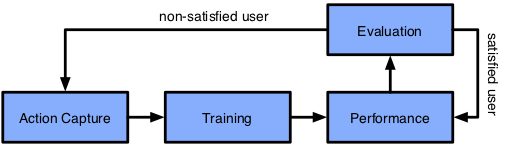}
\caption{The workflow of the presented method provides users the ability to directly and easily author the required computer puppetry scenarios through iterating the required steps.}
\label{fig5}
\end{figure}

A simple workflow (see Figure \ref{fig5}) allows users to interact directly with the MOP framework. The typical workflow is split into three basic parts, as follows:
\begin{itemize}
\item \textbf{Action Capture:} The user captures examples of actions that should be used to control the motions of the virtual character.
\item \textbf{Training:} The HMM, which is the core component of the MOP framework, is trained to recognize the user's actions. 
\item \textbf{Performance:} The user can use his/her body to animate the virtual character. 
\item \textbf{Evaluation:} In this case, the user evaluates the results. Depending on the satisfaction of the user, a captured action might be replaced by a new action, or the user might need to adjust the window size that is used to recognize the input actions. This process continues until the user becomes satisfied with the displayed results based on his/her performance.
\end{itemize}

\subsubsection{Examples}
\label{sec412}
A variety of examples are given to illustrate the efficiency and flexibility of the MOP framework. The video that accompanies this paper illustrates all of the examples presented in this section. 

\begin{figure*}[htb]
\centering
\includegraphics[width=1\textwidth]{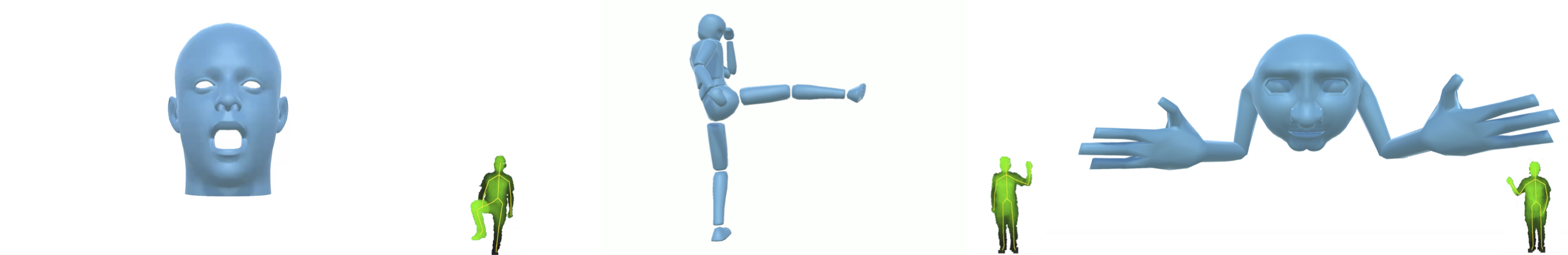}
\caption{A user controls a face model (right), a human-like character (middle), and a virtual creature (left) using his body.}
\label{fig6}
\end{figure*}

\begin{figure}[htb]
\centering
\includegraphics[width=0.95\columnwidth]{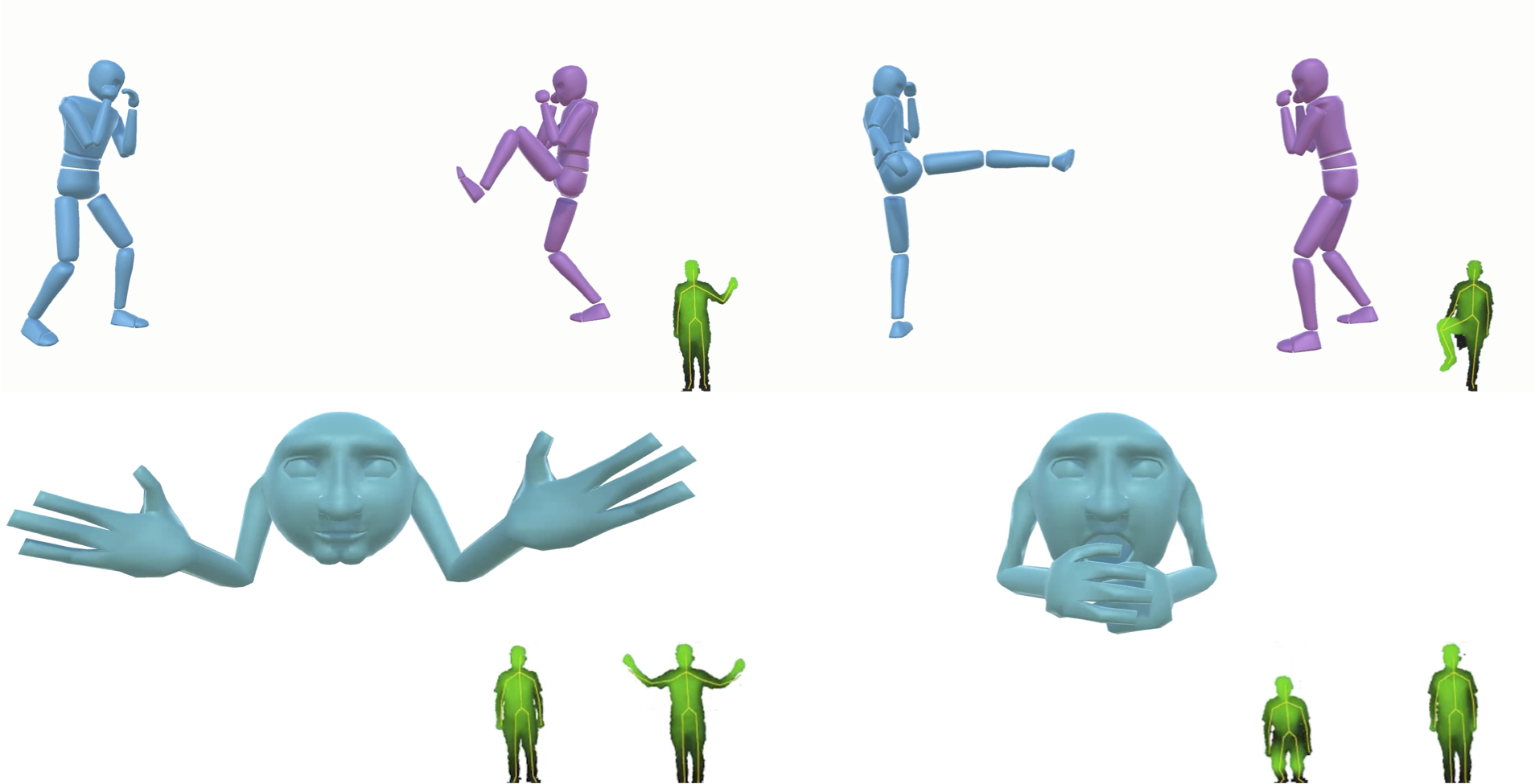}
\caption{A user controls two characters (top) using his body, and two users control a virtual creature (bottom) using their bodies.}
\label{fig7}
\end{figure}

\begin{figure}[htb]
\centering
\includegraphics[width=1\columnwidth]{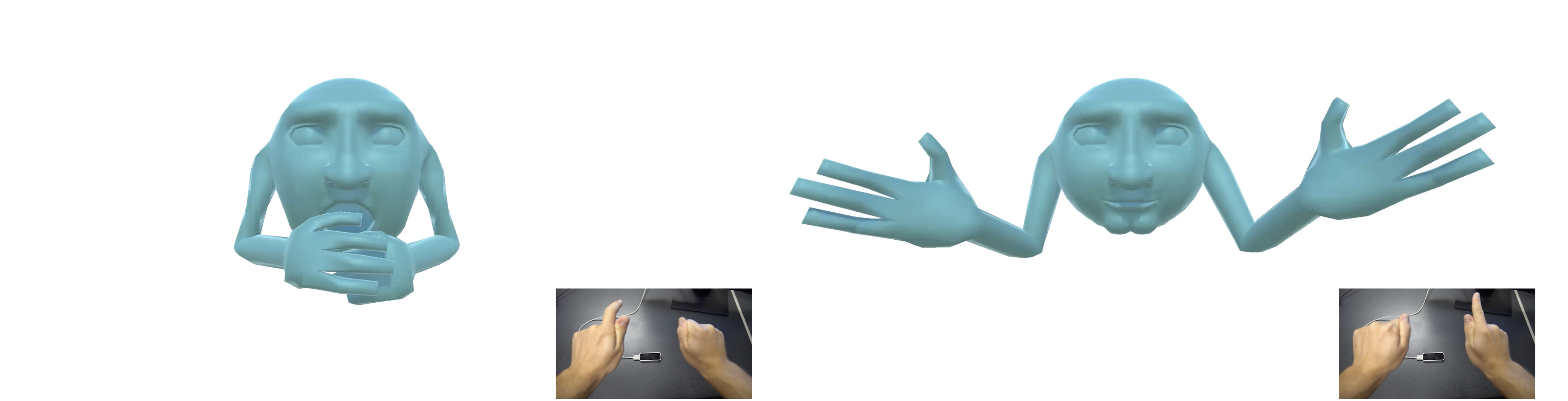}
\caption{A user controls a virtual creature using finger gestures from both hands.}
\label{fig8}
\end{figure}

\begin{figure}[htb]
\centering
\includegraphics[width=1\columnwidth]{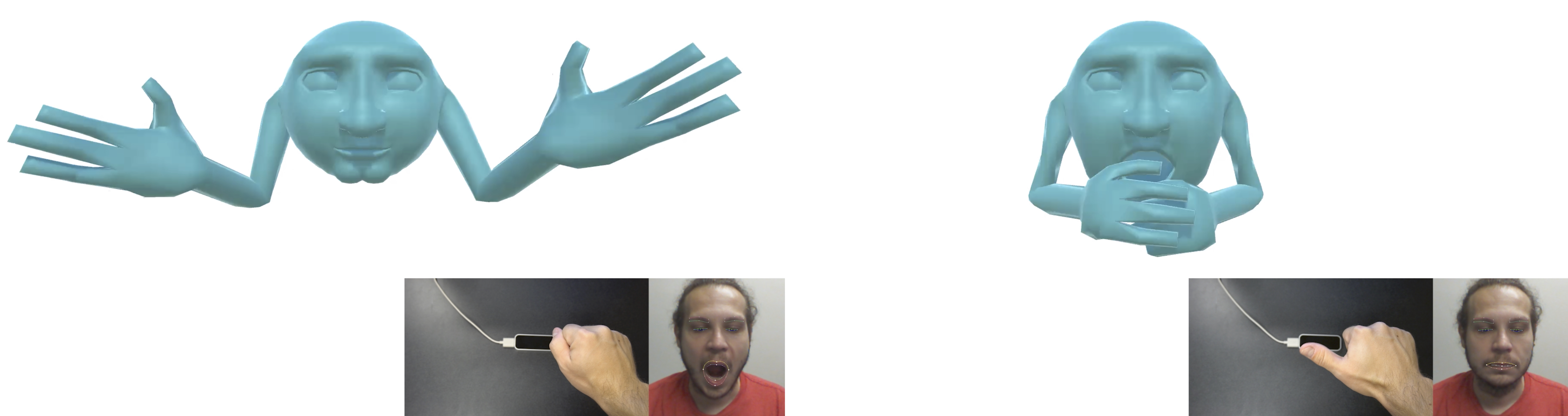}
\caption{A user controls a virtual creature using his face and fingers.}
\label{fig9}
\end{figure}

Figure \ref{fig6} illustrates a user controlling the expressions of a face model, the full-body motions of the human character, and a virtual creature using his body. Figure \ref{fig7} illustrates a single user controlling two characters using his body and also two users controlling a single character using their bodies. In the previously mentioned examples, the full-body motion of the user is captured using the Kinect motion capture sensor. Use of the Leap motion controller provides users the ability to control the virtual creature by using the fingers of both hands (see Figure \ref{fig8}). MOP also allows users to assign multiple input sources to a single character. Given this functionality, a final example is presented in which the user controls the motion of a virtual creature using his face and fingers. In this example, the face of the user is captured using the Kinect motion capture device, and the user's fingers are captured using the Leap sensor. Figure \ref{fig9} illustrates this example.

Here the following should be noted. When the user controls more than one character, the HMM model is trained independently for the user's action that should be assigned to each character's motion; thus, the user can control two characters simultaneously. When two users or more are responsible for the motion of a single character, the puppetry author should assign the actions of each user to the corresponding motions of the character. It should be noted that the actions of the users are captured independently although they are treated as a combined entry. Finally, in cases where two users or more are responsible for controlling the motion of multiple characters, the actions of the users are captured independently, and they are treated independently as separate entries.

\subsection{Evaluation}
\label{sec42}
The following sub-sections evaluate MOP in terms of performance as well as against that of existing approaches.

\subsubsection{Performance Evaluation}
\label{sec421}
The evaluation process examines the performance of the MOP framework based on different character control scenarios. Table \ref{tab1} presents the results of the evaluation process, which are based on an Intel i7-6700 CPU at 3.4 GHz with 16GB of RAM and an NVIDIA GeForce GTX 1060 with 6GB of RAM.

The presented methodology works in real time in all the examined character puppetry scenarios. Depending on the scenario, both the frames per second (FPS) and latency change. The results' values are all quite close to each other, and the performance of MOP remains on interactive framerates. The average framerate of the examined character control scenarios is approximately 84.45 FPS, and the latency is approximately 11.63 ms. The results suggest that this framework can be used in a variety of applications that are related to computer puppetry and motion control requiring run-time efficiency.

\begin{table*}[]
\centering
\begin{tabular}{|p{3.5cm}|p{1.3cm}|p{2cm}|p{1.8cm}|p{1cm}|l|l|l|}
\hline
~                                                 & \textbf{No. of Motions} & \textbf{No. of Characters} & \textbf{No. of Input Sources} & \textbf{No. of Users} & \textbf{Sensor} & \textbf{Latency} & \textbf{FPS} \\ \hline
\multirow{2}{*}{\textbf{Skeletal Animation}} & 5                       & 1                          & 1                             & 1                                                               & Kinect           & 7 ms             & 89           \\ \cline{2-8} 
                                                                                       & 10                      & 1                          & 1                             & 1                                                               & Kinect           & 9 ms             & 88           \\ \hline
\multirow{2}{*}{\textbf{Non-Skeletal Animation}}                                       & 5                       & 1                          & 1                             & 1                                                               & Kinect           & 6 ms             & 90           \\ \cline{2-8} 
                                                                                       & 10                      & 1                          & 1                             & 1                                                               & Leap             & 9 ms             & 86           \\ \hline
\multirow{2}{*}{\textbf{Hybrid Animation}}                                             & 5                       & 1                          & 1                             & 1                                                               & Kinect           & 6 ms             & 91           \\ \cline{2-8} 
                                                                                       & 10                      & 1                          & 2                             & 2                                                               & Kinect \& Leap   & 15 ms            & 81           \\ \hline
\multirow{3}{*}{\textbf{Multiple Characters}}                                          & 10                      & 2                          & 1                             & 1                                                               & Kinect           & 11 ms            & 85           \\ \cline{2-8} 
                                                                                       & 15                      & 3                          & 2                             & 2                                                               & Kinect \& Leap   & 17 ms            & 79           \\ \cline{2-8} 
                                                                                       & 20                      & 4                          & 2                             & 2                                                               & Kinect \& Leap   & 21 ms            & 76           \\ \hline
\multirow{2}{*}{\textbf{Multiple Users}}                                               & 10                      & 2                          & 1                             & 2                                                               & Kinect           & 11 ms            & 83           \\ \cline{2-8} 
                                                                                       & 10                      & 2                          & 2                             & 2                                                               & Kinect \& Leap   & 16 ms            & 81           \\ \hline
\multicolumn{6}{|r|}{\textbf{Average}}                                                                                                                                                                                                                             & 11.36 ms           & 84.45        \\ \hline
\end{tabular}
\caption{The performance of the MOP framework based on a variety of example scenarios.}
\label{tab1}
\end{table*}

\subsubsection{Comparison with Previous Methods}
\label{sec422}
A comparison was conducted to illustrate the differences between the MOP framework and previous character animation approaches. Table \ref{tab2} illustrates several criteria and functionalities that are divided into 12 categories, which were collected to compare MOP with other character animation frameworks and computer puppetry methods. 

\begin{table}[tb]
\scriptsize%
\centering%
\begin{tabu}{r*{7}{c}*{2}{r}}
\toprule
~ & \rotatebox{90}{Dontcheva et al. \cite{ref46}} & \rotatebox{90}{Ishigaki et al. \cite{ref20}} & \rotatebox{90}{Yamane et al. \cite{ref49}} & \rotatebox{90}{Seol et al. \cite{ref47}} & \rotatebox{90}{Rhodin et al. \cite{ref18}} & \rotatebox{90}{Chen et at al. \cite{ref19}} & \rotatebox{90}{MOP}\\
\hline
& \multicolumn{7}{c}{\textbf{Input Data}} \\
\hline
Full-Body		& ~ & \checkmark & \checkmark & \checkmark & \checkmark & \checkmark & \checkmark\\
Fingers 		& ~ &~ & ~ & ~ & \checkmark & ~ &\checkmark\\
Face 		& ~ &  ~ & ~ & ~ & \checkmark & ~ &  \checkmark  \\
Other 		& \checkmark &  ~ & ~ & ~ & ~ & ~ & ~ \\
    \hline
  & \multicolumn{7}{c}{\textbf{Number of Users}}\\
  \hline
  Single User 		& \checkmark &  \checkmark & ~ & \checkmark & \checkmark & ~ & \checkmark\\
  Multiple Users 	& ~ &  ~ & ~ & ~ & ~ & \checkmark & \checkmark\\
    \hline
  & \multicolumn{7}{c}{\textbf{Input Sensors}} \\
  \hline
  Single Sensor 	& \checkmark &  \checkmark & ~ & \checkmark & ~ & \checkmark & \checkmark\\
  Multiple Sensors 	& ~ &  ~ & ~ & ~ & \checkmark & ~ & \checkmark\\
    \hline
  & \multicolumn{7}{c}{\textbf{Input Sources}} \\
  \hline
  Single Source 	& \checkmark &  \checkmark & ~ & \checkmark & \checkmark & ~ & \checkmark\\
  Multiple Sources 	& ~ &  ~ & ~ & ~ & ~ & \checkmark & \checkmark   \\
    \hline
  & \multicolumn{7}{c}{\textbf{Motion Representation}} \\
    \hline
  Skeletal Animation 	& \checkmark &  \checkmark & \checkmark & \checkmark & \checkmark & ~ & \checkmark   \\
  Blendshapes	 	& ~ &  ~ & ~ & ~ & \checkmark & \checkmark & \checkmark   \\
  Hybrid (Skeletal \& Blendshapes)	& ~ &  ~ & ~ & ~ & ~ & ~ & \checkmark   \\
  \hline
  & \multicolumn{7}{c}{\textbf{Target Character}} \\
  \hline
  Human-like Characters 	& \checkmark &  \checkmark & ~ & ~ & \checkmark & ~ & \checkmark   \\
  Non-human-like Characters 	& \checkmark &  ~ & \checkmark & \checkmark & \checkmark & \checkmark & \checkmark   \\
    \hline  
  & \multicolumn{7}{c}{\textbf{Number of Characters}} \\
  \hline
  Single Character 		& \checkmark &  \checkmark & \checkmark & \checkmark & \checkmark & \checkmark & \checkmark   \\
  Multiple Characters 	& ~ &  ~ & ~ & ~ & ~ & ~ & \checkmark   \\
    \hline  
  & \multicolumn{7}{c}{\textbf{Character Animation}} \\
  \hline
  Motion Synthesis 		& \checkmark &  \checkmark & \checkmark & \checkmark & \checkmark & \checkmark & ~   \\
  Motion Diisplay 		& ~ &  ~ & ~ & ~ & ~ & ~ & \checkmark   \\
    \hline  
  & \multicolumn{7}{c}{\textbf{Performance Capture }} \\
  \hline
  Joint Manipulation 		& ~ &  \checkmark & ~ & ~ & ~ & ~ & ~   \\
  Activity Recognition 	& \checkmark &  \checkmark & ~ & \checkmark & \checkmark & \checkmark & ~  \\
  Activity Follower 		& ~ &  ~ & ~ & ~ & ~ & ~ & \checkmark   \\
    \hline      
  & \multicolumn{7}{c}{\textbf{System Performance}} \\
  \hline
  Real-Time 	& \checkmark &  \checkmark & ~ & \checkmark & \checkmark & \checkmark & \checkmark  \\
    \hline        
  & \multicolumn{7}{c}{\textbf{Naturalness}} \\
  \hline
  Natural-Looking Motion 	& \checkmark &  \checkmark & \checkmark & \checkmark & \checkmark & \checkmark & \checkmark  \\
    \hline            
  & \multicolumn{7}{c}{\textbf{Dynamics}} \\
  \hline
  Physics-Based Animation 	& ~ &  \checkmark & ~ & ~ & ~ & ~ & ~  \\
  \bottomrule
  \end{tabu}%
   \caption{Comparing previously developed computer puppetry approaches with the MOP framework.}
  \label{tab2}
\end{table}

The first category, input data, examines the user's body part inputs, which can be used as control parameters to animate the virtual character. This category examines the possibility of using inputs retrieved from the full-body, fingers, face, and other sources. The second category, number of users, examines the possibility of the methods to provide functionality for multiple users. The third category, input sensors, examines the possibility of using multiple sensors; and the fourth category, input sources, examines the possibility of using single or multiple input sources (e.g., a combination of face and fingers). The fifth category, motion representation, examines the representation that a motion requires to animate a character. The sixth category, target character, examines the system's ability to handle human-like or non-human-like characters. The seventh category, number of characters, examines the possibility of animating multiple characters simultaneously. The eighth category, character animation, examines the possibility of synthesizing animation or simply displaying the animation of the virtual character. The ninth category, performance capture, examines how the user's motion is used to animate a character. Finally, the last three categories - system performance, naturalness, and dynamics - examine the possibility of the systems to work in real time, the naturalness of the final motion, and the use of physics-based animation, respectively.

The basic concept behind MOP is the need for easy-to-use, fast, flexible performance-driven computer puppetry authoring frameworks. By comparing MOP with the methods presented in Table \ref{tab2}, it can be easily stated that many functionalities can be integrated or considered to design a powerful computer puppetry authoring framework. Moreover, each of the existing studies has advantages and disadvantages. Finally, the comparison is based on the basic functionalities that each of the approaches provides. Even if the methods could be extended to use multiple sources or increase the number of users that could interact with the content, the presented evaluation is based on what the authors presented and discussed in the corresponding papers.

The solutions that are closest to MOP in terms of translating the user's activity to that of the virtual character are those proposed by Rhodin et al. \cite{ref18}, Ishigaki et al. \cite{ref20}, Chen et at al. \cite{ref19}, and Seol et al. \cite{ref47}. In terms of animating a variety of virtual characters (both humanoid and non-humanoid), the methods proposed by Rhodin et al. \cite{ref18} and Seol et al. \cite{ref47} can be characterized as similar to MOP. In regard to using multiple and different input sensors and sources, only the methods proposed by Rhodin et al. \cite{ref18} can be considered similar to the MOP framework. In terms of the performance capture process, Ishigaki et al. \cite{ref20} used activity recognition and direct joint manipulation to animate the virtual character. When examining the way that the motion of a character is displayed, it is obvious that only MOP allows the motion itself to evolve in time based on the activity-following method compared with the other approaches that use motion synthesis techniques.

Compared with Dontcheva et al. \cite{ref46}, a mapping process allows a user to animate a character in a natural-looking way. However, it can be stated that even if the character is able to follow the movements of the controller for each layer separately, the need for non-commercial specialized equipment for controlling a virtual character can be considered a disadvantage. The approach presented by Yamane et al. \cite{ref49} provides the ability to animate a number of different non-human characters in a natural-looking way; however, the motion is synthesized off-line and does not provide the user the freedom to control the virtual character. It is also important to note that the system presented by Yamane et al. \cite{ref49} is not a real-time performance-driven animation. 
Generally, the examined frameworks and methods cover only a limited number of functionalities and lack the generalization that is provided by the MOP framework. Moreover, none of the examined approaches provides the activity-following functionality for controlling the evolution of a virtual character's motion. Similarly, functionalities such as the control of hybrid motions (motions represented by skeletal keyframes and blendshape-based keyframes simultaneously), the simultaneous control of two characters, or the simultaneous use of multiple input sources are not provided by any of the examined approaches.

\section{Conclusions and Future Work}
\label{sec5}

This paper introduces an animation-by-demonstration computer puppetry authoring framework. The real-time character control is achieved using HMM to train and predict the evolution of the user's actions. By assigning the character controller mechanism to the forward algorithm, it is possible to predict the activity of the user that corresponds to the motion of the character that should be displayed as well as the evolution of the character's motion based on the evolution of the user's action. The development and demonstration of a variety of examples shows the flexibility of the presented methodology in handling diverse characters, inputs, and sources, thereby providing users the ability to control the required animation quite efficiently for various computer puppetry scenarios.

In its current form, the developed algorithm is able to handle a variety of different user actions. Since the recognition process takes into account the user's total number of captured joints, it is easy to predict the input action and display the necessary motion of the character for different body parts of the user. However, there are cases where the user might require similar actions to control the virtual character. An example includes actions where the user employs only his/her right hand. In this case, the system can recognize different actions; however, motions can be recognized incorrectly if they are too similar. We believe that such wrong estimations might affect the way the user controls the virtual character. Therefore, a future improvement of the MOP framework could provide feedback for users immediately after recording the action, indicating whether the new action would affect the recognition process. Additionally, we also believe that improving the recognition to generalize forms of user actions would also benefit the MOP framework, which means that users would be able to control the virtual character even if an action is performed with a variation of the captured one. 

The character animation framework could be improved in additional ways besides the recognition process. For example, it would be interesting to integrate additional character control mechanisms that would enhance the flexibility and applicability of the current version. The addition of physics-related reactions \cite{ref16}\cite{ref20} could benefit the range of motion, which could be synthesized by the current versions. Other functionalities that would improve the naturalness of the character's motion should also be considered. A simple example includes the relationship description \cite{ref9}, which would allow the developed framework to handle interactions between multiple characters more efficiently. Finally, the extension of the current HMM to a hierarchical model \cite{ref35} with reactive interpolations \cite{ref53} is another improvement that would benefit the MOP framework, since it would predict the progression of the character's motion while synthesizing the motion sequences, which is a functionality that is not provided in the current version of the MOP framework.


{\small
\bibliographystyle{unsrt}
\bibliography{cvmbib}

\begin{thebibliography}{10}

\bibitem{ref37}
N.~Sarris and M.~G. Strintzis.
\newblock {\em 3D modeling and animation: Synthesis and analysis techniques for
  the human body}.
\newblock IGI Global, 2005.

\bibitem{ref26}
M.~Gleicher.
\newblock Retargetting motion to new characters.
\newblock In {\em Annual Conference on Computer Graphics and Interactive
  Techniques}, pages 33--42. ACM, July 1998.

\bibitem{ref38}
J.~McCann and N.~Pollard.
\newblock Responsive characters from motion fragments.
\newblock {\em ACM Transactions on Graphics}, 26(3):Article No. 6, August 2007.

\bibitem{ref2}
M.~Oshita.
\newblock Generating animation from natural language texts and semantic
  analysis for motion search and scheduling.
\newblock {\em The Visual Computer}, 26(5):339--352, 2010.

\bibitem{ref7}
C.~Mousas and C.-N. Anagnostopoulos.
\newblock Chase: character animation scripting environment.
\newblock In {\em ACM SIGGRAPH International Conference on Virtual Reality
  Continuum and its Applications in Industry}, pages 55--62, 2015.

\bibitem{ref11}
S.~Levine, P.~Kr{\"a}henb{\"u}hl, S.~Thrun, and V.~Koltun.
\newblock Gesture controllers.
\newblock {\em ACM Transactions on Graphics}, 29(4):Article No. 124, 2010.

\bibitem{ref40}
J.~Davis, M.~Agrawala, E.~Chuang, Z.~Popovi{\'c}, and D.~D.~Salesin.
\newblock A sketching interface for articulated figure animation.
\newblock In {\em ACM SIGGRAPH/Eurographics Symposium on Computer Animation},
  pages 320--328. Eurographics Association, July 2003.

\bibitem{ref18}
H.~Rhodin, J.~Tompkin, K~{In Kim}, K.~Varanasi, H.~P. Seidel, and C.~Theobalt.
\newblock Interactive motion mapping for real‐time character control.
\newblock {\em Computer Graphics Forum}, 33(2):273--282, May 2014.

\bibitem{ref19}
J.~Chen, S.~Izadi, and A.~Fitzgibbon.
\newblock Kin{\^e}tre: animating the world with the human body.
\newblock In {\em ACM Symposium on User Interface Software and Technology},
  pages 435--444. ACM, October 2012.

\bibitem{ref6}
C.~Ouzounis, C.~Mousas, C.-N. Anagnostopoulos, and P.~Newbury.
\newblock Using personalized finger gestures for navigating virtual characters.
\newblock In {\em Workshop on Virtual Reality Interaction and Physical
  Simulation}, pages 5--14, 2015.

\bibitem{ref12}
W.-C. Lam, F.~Zou, and T.~Komura.
\newblock Motion editing with data glove.
\newblock In {\em ACM SIGCHI International Conference on Advances in Computer
  Entertainment Technology}, pages 337--342, 2004.

\bibitem{ref13}
R.~Y. Wang and J.~Popovi{\'c}.
\newblock Real-time hand-tracking with a color glove.
\newblock {\em ACM Transactions on Graphics}, 28(3):Article No. 63, July 2009.

\bibitem{ref14}
T.~Shiratori and J.~K. Hodgins.
\newblock Accelerometer-based user interfaces for the control of a physically
  simulated character.
\newblock {\em ACM Transactions on Graphics}, 27(5):Article No. 123, 2008.

\bibitem{ref17}
R.~Slyper and J.~K. Hodgins.
\newblock Action capture with accelerometers.
\newblock In {\em ACM SIGGRAPH/Eurographics Symposium on Computer Animation},
  pages 193--199. Eurographics Association, July 2008.

\bibitem{ref41}
A.~Jacobson, D.~Panozzo, O.~Glauser, C.~Pradalier, O.~Hilliges, and
  O.~Sorkine-Hornung.
\newblock Tangible and modular input device for character articulation.
\newblock {\em ACM Transactions on Graphics}, 33(4):Article No. 82, 2014.

\bibitem{ref21}
T.~Mukai and S.~Kuriyama.
\newblock Geostatistical motion interpolation.
\newblock {\em ACM Transactions on Graphics}, 24(3):1062--1070, July 2005.

\bibitem{ref22}
L.~Kovar and M.~Gleicher.
\newblock Flexible automatic motion blending with registration curves.
\newblock In {\em ACM SIGGRAPH/Eurographics Symposium on Computer Animation},
  pages 214--224. Eurographics Association, July 2003.

\bibitem{ref24}
B.~van Basten and A.~Egges.
\newblock Motion transplantation techniques: a survey.
\newblock {\em IEEE Computer Graphics and Applications}, 32(3):16--23, 2012.

\bibitem{ref1}
C.~Mousas, P.~Newbury, and C.-N. Anagnostopoulos.
\newblock Splicing of concurrent upper-body motion spaces with locomotion.
\newblock In {\em Procedia Computer Science}, volume~25, pages 348--359, 2013.

\bibitem{ref25}
A.~Witkin and Z.~Popovic.
\newblock Motion warping.
\newblock In {\em Annual Conference on Computer Graphics and Interactive
  Techniques}, pages 105--108. ACM, September 1995.

\bibitem{ref23}
P.~Glardon, R.~Boulic, and D.~Thalmann.
\newblock Pca-based walking engine using motion capture data.
\newblock In {\em Computer Graphics International}, pages 292--298. IEEE, June
  2004.

\bibitem{ref8}
J.~Song, B.~Choi, Y.~Seol, and J.~Noh.
\newblock Characteristic facial retargeting.
\newblock {\em Computer Animation and Virtual Worlds}, 22(2‐3):187--194,
  2011.

\bibitem{ref15}
C.~Ouzounis, A.~Kilias, and C.~Mousas.
\newblock Kernel projection of latent structures regression for facial
  animation retargeting.
\newblock In {\em EUROGRAPHICS Workshop on Virtual Reality Interaction and
  Physical Simulation}, pages 59--65, 2017.

\bibitem{ref44}
T.~J. Cashman and K.~Hormann.
\newblock A continuous, editable representation for deforming mesh sequences
  with separate signals for time, pose and shape.
\newblock {\em Computer Graphics Forum}, 31(2):735--744, May 2012.

\bibitem{ref43}
S.~Levine, J.~M. Wang, A.~Haraux, Z.~Popovi{\'c}, and V~Koltun.
\newblock Continuous character control with low-dimensional embeddings.
\newblock {\em ACM Transactions on Graphics}, 31(4):Article No. 28, 2012.

\bibitem{ref28}
R.~Slyper and J.~K. Hodgins.
\newblock Action capture with accelerometers.
\newblock In {\em ACM SIGGRAPH/Eurographics Symposium on Computer Animation},
  pages 193--199. Eurographics Association, July 2008.

\bibitem{ref30}
X.~Wei, P.~Zhang, and J.~Chai.
\newblock Accurate realtime full-body motion capture using a single depth
  camera.
\newblock {\em ACM Transactions on Graphics}, 31(6):Article No. 188, 2012.

\bibitem{ref32}
D.~Raunhardt and R.~Boulic.
\newblock Immersive singularity‐free full‐body interactions with reduced
  marker set.
\newblock {\em Computer Animation and Virtual Worlds}, 22(5):407--419, 2011.

\bibitem{ref34}
H.~Liu, X.~Wei, J.~Chai, I.~Ha, and T.~Rhee.
\newblock Realtime human motion control with a small number of inertial
  sensors.
\newblock In {\em Symposium on Interactive 3D Graphics and Games}, pages
  133--140. ACM, February 2011.

\bibitem{ref4}
C.~Mousas, P.~Newbury, and C.-N. Anagnostopoulos.
\newblock Evaluating the covariance matrix constraints for data-driven
  statistical human motion reconstruction.
\newblock In {\em Spring Conference on Computer Graphics}, pages 99--106. ACM,
  2014.

\bibitem{ref10}
H.~Eom, D.~Choi, and J.~Noh.
\newblock Data‐driven reconstruction of human locomotion using a single
  smartphone.
\newblock {\em Computer Graphics Forum}, 33(7):11--19, 2014.

\bibitem{ref36}
J.~Min and J.~Chai.
\newblock Motion graphs++: a compact generative model for semantic motion
  analysis and synthesis.
\newblock {\em ACM Transactions on Graphics}, 31(6):Article No. 153, 2012.

\bibitem{ref45}
D.~J. Sturman.
\newblock Computer puppetry.
\newblock {\em IEEE Computer Graphics and Applications}, 18(1):38--45, 1998.

\bibitem{ref46}
M.~Dontcheva, G.~Yngve, and Z.~Popovi{\'c}.
\newblock Layered acting for character animation.
\newblock {\em ACM Transactions on Graphic}, 22(3):409--416, 2003.

\bibitem{ref47}
Y.~Seol, C.~O'Sullivan, and J.~Lee.
\newblock Creature features: online motion puppetry for non-human characters.
\newblock In {\em ACM SIGGRAPH/Eurographics Symposium on Computer Animation},
  pages 213--221, July 2013.

\bibitem{ref20}
S.~Ishigaki, T.~White, V.~B. Zordan, and C.~K. Liu.
\newblock Performance-based control interface for character animation.
\newblock {\em ACM Transactions on Graphics}, 28(3):Article No. 61, July 2009.

\bibitem{ref49}
K.~Yamane, Y.~Ariki, and J.~Hodgins.
\newblock Animating non-humanoid characters with human motion data.
\newblock In {\em ACM SIGGRAPH/Eurographics Symposium on Computer Animation},
  pages 169--178. Eurographics Association, July 2010.

\bibitem{ref31}
C.~Mousas and C.-N. Anagnostopoulos.
\newblock Performance-driven hybrid full-body character control for navigation
  and interaction in virtual environments.
\newblock {\em 3D Research}, 8(2):Article No. 18, 2017.

\bibitem{ref52}
T.~Otsuka, K.~Nakadai, T.~Ogata, and H.~G. Okuno.
\newblock Incremental bayesian audio-to-score alignment with flexible harmonic
  structure models.
\newblock In {\em ISMIR}, pages 525--530, 2011.

\bibitem{ref33}
J.~Fran{\c c}oise, B.~Caramiaux, and F.~Bevilacqua.
\newblock A hierarchical approach for the design of gesture-to-sound mappings.
\newblock In {\em Sound and Music Computing Conference}, pages 233--240, July
  2012.

\bibitem{ref39}
F.~Bettens~T. Todoroff.
\newblock Real-time dtw-based gesture recognition external object for max/msp
  and puredata.
\newblock In {\em Sound and Music Computing Conference}, pages 30--35, July
  2009.

\bibitem{ref50}
F.~Bevilacqua, B.~Zamborlin, A.~Sypniewski, N.~Schnell, F.~Gu{\'e}dy, and
  N.~Rasamimanana.
\newblock Continuous realtime gesture following and recognition.
\newblock In {\em Gesture in embodied communication and human-computer
  interaction}, pages 73--84. Springer Berlin Heidelberg, 2010.

\bibitem{ref48}
S.~S. Fels and G.~E. Hinton.
\newblock Glove-talk: A neural network interface between a data-glove and a
  speech synthesizer.
\newblock {\em IEEE Transactions on Neural Networks}, 4(1):2--8, 1993.

\bibitem{ref29}
R.~Fiebrink, P.~R. Cook, and D.~Trueman.
\newblock Human model evaluation in interactive supervised learning.
\newblock In {\em SIGCHI Conference on Human Factors in Computing Systems},
  pages 147--156. ACM, May 2011.

\bibitem{ref27}
A.~Mori, S.~Uchida, R.~Kurazume, R.~I. Taniguchi, T.~Hasegawa, and H.~Sakoe.
\newblock Early recognition and prediction of gestures.
\newblock In {\em International Conference on Pattern Recognition}, pages
  560--563. IEEE, August 2006.

\bibitem{ref42}
D.~J. Berndt and J.~Clifford.
\newblock Using dynamic time warping to find patterns in time series.
\newblock In {\em KDD workshop}, pages 359--370, 1994.

\bibitem{ref51}
L.~Rabiner.
\newblock A tutorial on hidden markov models and selected applications in
  speech recognition.
\newblock {\em Proceedings of the IEEE}, 77(2):257--286, 1989.

\bibitem{ref5}
{Microsoft}.
\newblock {Kinect SDK} from
  \url{https://www.microsoft.com/en-us/kinectforwindows/}, 2017.

\bibitem{ref3}
{Leap Motion}.
\newblock {Developers SDK} from \url{https://developer.leapmotion.com/}, 2017.

\bibitem{ref16}
X.~Liang, L.~Hoyet, W.~Geng, and F.~Multon.
\newblock Responsive action generation by physically-based motion retrieval and
  adaptation.
\newblock In {\em Motion in Games}, pages 313--324. Springer Berlin Heidelberg,
  2010.

\bibitem{ref9}
R.~A. Al-Asqhar, T.~Komura, and M.~G. Choi.
\newblock Relationship descriptors for interactive motion adaptation.
\newblock In {\em ACM SIGGRAPH/Eurographics Symposium on Computer Animation},
  pages 45--53, 2013.

\bibitem{ref35}
C.~Mousas and C.-N. Anagnostopoulos.
\newblock Real-time performance-driven finger motion synthesis.
\newblock {\em Computers \& Graphics}, 65:1--11, 2017.

\bibitem{ref53}
C.~Mousas.
\newblock Full-body locomotion reconstruction of virtual characters using a
  single inertial measurement unit.
\newblock {\em Sensors}, 17(11):Article No. 2589, November 2017.

\end{thebibliography}
}

\end{document}